\documentclass[twocolumn,10pt,final]{asme2e}
\usepackage{graphicx, bm}
\usepackage{amsmath}
\usepackage[printonlyused,nolist]{acronym}
\usepackage{url}
\usepackage{enumerate}
\usepackage{multirow}
 
\confshortname{IDETC/CIE 2013}
\conffullname{the 2013 ASME International Design Engineering Technical Conferences \& Computers and Information in Engineering Conference}

\confdate{4-7}
\confmonth{August}
\confyear{2013}
\confcity{Portland, Oregon}
\confcountry{USA}

\papernum{DETC2013-12188}

\title{The Failure Tolerance of Mechatronic Software Systems to Random and Targeted Attacks}

\author{Dharshana Kasthurirathna    
    \affiliation{
	Centre for Complex Systems Research\\
	Faculty of Engineering \& Information Technologies\\
	University of Sydney\\
	Sydney, NSW, 2006, Australia \\
	dkas2394@uni.sydney.edu.au
    }
}

\author{Andy Dong \thanks{\vspace{-1em}Address all correspondence to this author.}    
    \affiliation{
	Faculty of Engineering \& Information Technologies\\
	University of Sydney\\
	Sydney, NSW, 2006, Australia \\
	andy.dong@sydney.edu.au 
    }
}

\author{Mahendrarajah Piraveenan   
    \affiliation{
	Centre for Complex Systems Research\\
	Faculty of Engineering \& Information Technologies\\
	University of Sydney\\
	Sydney, NSW, 2006, Australia \\
	mahendrarajah.piraveenan@sydney.edu.au 
    }
}

\author{Irem Y. Tumer   
    \affiliation{
	Complex Engineered Systems Design (CESD) Laboratory\\
School of Mechanical, Industrial, \& Manufacturing Engineering\\
Oregon State University\\
Corvallis, OR 97331-6001\\
irem.tumer@oregonstate.edu}
}

\begin{document}

\maketitle    

\begin{abstract}
{\it This paper describes a complex networks approach to study the failure tolerance of mechatronic software systems under various types of hardware and/or software failures. We produce synthetic system architectures based on evidence of modular and hierarchical modular product architectures and known motifs for the interconnection of physical components to software. The system architectures are then subject to various forms of attack. The attacks simulate failure of critical hardware or software. Four types of attack are investigated: degree centrality, betweenness centrality, closeness centrality and random attack. Failure tolerance of the system is measured by a `robustness coefficient', a topological `size' metric of the connectedness of the attacked network. We find that the betweenness centrality attack results in the most significant reduction in the robustness coefficient, confirming betweenness centrality, rather than the number of connections (i.e. degree), as the most conservative metric of component importance. A counter-intuitive finding is that ``designed" system architectures, including a bus, ring, and star architecture, are not significantly more failure-tolerant than interconnections with no prescribed architecture, that is, a random architecture. Our research provides a data-driven approach to engineer the architecture of mechatronic software systems for failure tolerance.}
\end{abstract}


\section{Introduction}
While mechatronic systems, that is, mechanical systems controlled by or having software embedded in them, have been in operation for quite some time, the understanding of the failure properties of integrated mechatronic software systems has not kept up with their adoption. The unexplained failure of these systems, including heretofore recently unexplained ones such as the sudden uncommanded pitch-down manoeuvres of a Qantas Airbus A330 over Western Australia in 2008, have raised the priority of understanding the complex interactions between hardware and software in mechatronic software systems. Of concern in this paper is the performance of the overall mechatronic software system due to failures in hardware components or software code.

As engineered mechatronic systems become more architecturally complex, which is to say, have an increasingly complex physical product architecture~\cite{Simon1962} and a complex set of inter-relations with software code, the study of the tolerance of these systems to failures of physical components and software becomes increasingly challenging. Abstracting such systems from the viewpoint of energy flow between components seems to be the preferred method of modelling and analysis.

Design methods such as the Function-Failure Design Method~\cite{Stone2005} and the \ac{FFIP} analysis framework~\cite{papakonstantinou:031007,Kurtoglu2010,Kurtoglu2008} have begun to adopt graph-based approaches. Methods for hazard identification such as the \ac{MFIP} framework similarly identify components likely to fail when a hazard is propagated from one component to another~\cite{Mehrpouyan2012}. The basic idea underlying these approaches is to model the function of components and the flow of energy between them as a network, and then to identify the failure propagation through the network. We extend this basic idea to study mechatronic software systems as interdependent networks~\cite{Brummitt20032012,PhysRevLett.105.048701,Buldyrev2010}, wherein the physical architecture of the hardware and the software (code) architecture are modelled as interacting networks. We study the topological properties of interdependent mechatronic software networks associated with system tolerance to failure of components because the topology of complex networks is known to influence the behaviour of the network in response to the failure or modification of nodes~\cite{Albert2000}.

The importance of studying the failure of systems based upon the topology of its hardware and software architecture has become even more important due to recent observations that modular systems tend to have low robustness even under increasing robustness of individual components~\cite{2011arXiv1102.5085B}. Modular systems have high robustness only when the failure of components can be isolated to its module or when alternative (redundant) pathways between components can be designed in~\cite{Ash2007}. Designing a complex system as a single large module with a high density of connections is not always practical though. Conversely, the hierarchical modular structure of brain networks has been observed to enhance the brain's robustness, but too much modularity reduces its functionality~\cite{1367-2630-14-2-023005}, which is consistent with observations on the loss of performance in highly modular engineered systems~\cite{Holtta2005}. We cite these references against the backdrop of a preference for increasingly modular physical architectures in complex engineered products~\cite{Gershenson2003,Ishii2003} despite loss of performance improvement~\cite{Ethiraj2004} and making it easier for competitors to copy designs~\cite{Ethiraj2008}. Thus, the recommendation and preference for modular physical system architectures may have the unintended downside of making the systems less tolerant to failure.

In summary, although great efforts are taken to engineer reliable physical components and error-free software code, little is known about the compound effect of system connectivity and individual failures within each system on the mechatronic software system. Specifically, it is not known if the failure properties of the physical system due to the physical architecture and the software system due to its software architecture should be studied individually or whether the failure properties are, in a sense, an additive sum of the two architectures combined. To investigate the failure of a combined mechatronic software system, we take the product architecture of hardware and software architecture, study their individual failure properties, and then study the failure properties of the two interdependent networks using a topological metric known as a `robustness coefficient'~\cite{robustness,Piraveenan2012}.

\section{Background}
\label{background}

Until recently, the study of the failure of complex engineered systems had focused on the reliability of individual components and on simulating the performance of the system under various input conditions in order to understand system-wide effects of input variability~\cite[e.g.]{kim:121401,tu:557}. Extending from this base of reliability-based system design, engineering design researchers have begun to direct their efforts toward the robustness of complex engineered systems so as to engineer systems that are able to recover from disturbances and failures to continue functioning within the expected performance envelope~\cite{Youn2011}. At the core of these methods, though, is a focus on the component and the influence of the component on the failure of the system, or on the external conditions and their causal influence on the failure of the system vis-\`{a}-vis the components.

Recent research evidence emanating from the field of complex networks suggests that the vulnerability of a system is at least equally dependent upon the architecture of the system, which takes into account the connectivity of the components, as the components themselves \cite{Albert2000,costareview,doro,EPL,pir01}. Research in the field of complex networks has shown that not all network topologies display equivalent properties of failure tolerance to errors in nodes (components). In other words, some networks exhibit topological weakness. Scale-free networks, that is, networks wherein a few nodes have a high number of connections but most nodes have very few connections to other nodes, have a high degree of tolerance against the random failure of nodes~\cite{Albert2000}. Braha and Bar-Yam similarly find that the dynamic behaviour of inhomogenous networks is resilient to random failures of nodes but highly sensitive to modifications to essential nodes~\cite{Braha2007}. Similarly, systems containing hubs have been found to have higher levels of quality~\cite{Sosa2011, PPZ}, suggesting that systems having scale-free properties may be preferred over their small-world or exponential (homogenous) degree counterparts when it comes to tolerance to random failures or modifications to nodes. This failure tolerance is not true of other network topologies such as random networks or homogeneous networks, wherein most nodes have approximately the same number of links to other nodes. However, the scale-free networks also lose connectivity quickly when the most important nodes fail, which means that the network becomes fragmented into isolated nodes and communities of nodes. In an engineered system, this could be equivalent to total system failure, especially if the functional network is disconnected, since the system would no longer behave as designed, or if a critical module is separated from the rest of the system.

Mechatronic software systems represent a more interesting and qualitatively different type of network though, one which is more `naturally' represented as two interdependent networks, a hardware network (physical architecture) and a software network (software architecture). The hardware network is partially dependent upon the software network, which operates the control logic for the hardware. In turn, the software is partially dependent upon the hardware for power, sensing, and actuation. These dependencies can make the system vulnerable to an avalanche of failures because a failure in the hardware cascades and causes failure in the software, which in turn may produce erratic control signals that produce abnormal behaviour by the hardware and so on. Recent theoretical research on interdependent networks shows altogether different failure behaviours for interdependent networks than for single networks. Inhomogenous degree distributions increase the vulnerability of interdependent networks to random failure~\cite{Buldyrev2010}, meaning that the advantage of an inhomogeneous single network becomes a disadvantage for interdependent networks. Likewise, cascading failures occur in interdependent networks when nodes in each network mutually depend on nodes in other networks such that that failures can cascade back and forth between the networks~\cite{PhysRevLett.107.195701}. Contrary to intuition, adding more interconnectivity between networks can become detrimental to the system even when the additional interconnectivity is beneficial to an individual network~\cite{Brummitt20032012}.

Based upon these critical observations on the failure properties of interdependent layered networks, this paper addresses the failure properties of mechatronic software systems. We synthesise modular and hierarchical modular physical architectures hardware networks, which are frequently found in product architectures, and scale-free software architectures, based upon theoretical evidence of their presence~\cite{Valverde2002}. We create an interdependent layered hardware-software network based on various rules for their interconnection. We describe the failure properties of the networks based upon a `robustness coefficient'~\cite{robustness,Piraveenan2012} and show how the failure properties differ from the failure properties of individual networks.

\section{Research Method}
\label{method}

\subsection{Network synthesis}
The initial part of the research focuses on a producing synthetic networks that model real-world product and software architectures. The synthetic hardware networks simulate a hardware system having modular and hierarchical modular product architectures, given the preference for modular product architectures in mechanical systems. To produce modular and hierarchical modular networks, we follow the method described by Sarkar and Dong~\cite{Sarkar2011}. This methodology involves `rewiring' each edge in a perfectly modular network to take away intra-community edges in each module with a rewiring probability $p$. By varying $p$, we obtain networks that have varying modularity levels.

Similarly, we produce synthetic, scale-free software networks. Software architecture have been demonstrated to show scale-free behaviour despite design strategies associated with software engineering preferring tree-like structures~\cite{Valverde2002}. Preferential attachment based generator models can be used to produce scale-free networks~\cite{Barabasi1999}. Therefore, we generated synthetic scale-free software networks based upon preferential attachment. The preferential attachment may represent software dependencies at different granularity levels such as class graphs or function graphs. These synthetic software networks will be connected to the hardware networks in the integrated hardware-software networks based upon various connection motifs.

The next part of the research focuses on producing synthetic hardware-software interdependent networks. We combine the hardware physical architecture network and software networks by integrating their adjacency matrices. An \textit{adjacency matrix} represents the edges between nodes in a network using a matrix representation. A binary adjacency matrix \textbf{A} represents whether or not there is an edge (physical architecture relationship) between nodes $i$ and $j$ as in 

\begin{equation}
A_{ij} = \left\{
\begin{array}{rl}
1 & \text{if an edge exists between nodes } i \text{ and } j\\
0 & \text{otherwise }
\end{array} \right.
\end{equation}

\begin{figure}
\includegraphics[width=4.05cm, height=4.05cm]{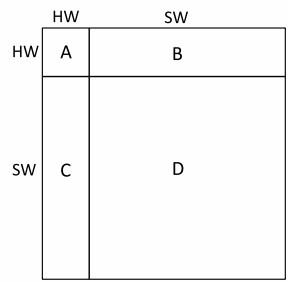}
\caption{\label{fig:F1}THE INTEGRATION OF THE ADJACENCY MATRICES OF MECHATRONIC SOFTWARE SYSTEM ARCHITECTURES}
\end{figure}

Fig.~\ref{fig:F1} shows how a typical mechatronic software system could be represented in this manner.

In a typical mechatronic software system, the hardware and the software system `interact' with each other to form an integrated interdependent network. As such, in our approach, we neither reduce the two networks into a single network nor do we study them separately. Instead, we integrate the two networks by considering the interactions between them as interdependent networks. Some hardware components depend upon the software for their behaviour and some software components depend upon hardware for input data.

In Fig.~\ref{fig:F1}, section A represents the hardware product architecture network, while D represents the software architecture network. Accordingly, the B and C adjacency matrix sections represent the interactions between the hardware components and software code, such as between a sensor (hardware) and the software code polling the sensor data. The fraction of non-zero values in B and C describe the fraction of nodes in A that depend upon D and the fraction of nodes in D that depend upon A, respectively. The total adjacency matrix could be regarded as a representation of the integrated hardware-software network in a mechatronic software system. This type of integration would minimise the loss of information that could occur when reducing the networks into a single network. At the same time, this approach would facilitate the studying of a mechatronic software system as a single system constituted by two separate hardware and software networks.

We investigate various ways in which the hardware-software networks could be interdependent. Network motifs have been used as an indicator of software architectural patterns. For example, given a network of software classes, the network motifs that may exist in that particular network would have a relationship to the architectural pattern or the rules that are used in developing the system~\cite{ValverSole-PRE05}. Extending on the same idea, we propose the usage of network motifs as an implication of the usage of a particular architecture in mechatronic software system design.

A network motif can be represented by a specific bit pattern in the hardware-software adjacency matrix. Therefore, by applying different adjacency patterns to the hardware--software adjacency matrix sections (namely B and C), we could manipulate the architecture of the hardware-software system that it resembles. Table~\ref{tab:T1} shows the adjacency patterns that could be used to emulate several well-known architectures in mechatronic software system design.

When generating the integrated adjacency matrix, we use the largest single block of each pattern that a mechatronic software system adjacency matrix section could accommodate.
 
\begin{table}[htbp]
	\centering
		 \small		
			 \caption{HW--SW ADJACENCY MATRIX BIT PATTERNS RESEMBLING DIFFERENT EMBEDDED SYSTEM ARCHITECTURES}
\begin{tabular}{crrrrrr}
	\hline
	Architecture & \multicolumn{6}{c}{Adjacency Pattern} \\
	\hline
	\multirow{6}{*}{Bus} & 0 & 1 & 0 & 0 & 1 & 0\\
 & 1 & 0 & 1 & 0 & 0 & 1\\
 & 0 & 1 & 0 & 1 & 0 & 0\\
 & 0 & 0 & 1 & 0 & 0 & 0\\
 & 1 & 0 & 0 & 0 & 0 & 0\\
 & 0 & 1 & 0 & 0 & 0 & 0 \\ \hline
 	\multirow{6}{*}{Ring} & 0 & 1 & 0 & 0 & 0 & 1\\
 & 1 & 0 & 1 & 0 & 0 & 0\\
 & 0 & 1 & 0 & 1 & 0 & 0\\
 & 0 & 0 & 1 & 0 & 1 & 0\\
 & 0 & 0 & 0 & 1 & 0 & 1\\
 & 1 & 0 & 0 & 0 & 1 & 0 \\ \hline
 	\multirow{6}{*}{Star} & 0 & 1 & 1 & 1 & 1 & 1\\
 & 1 & 0 & 0 & 0 & 0 & 0\\
 & 1 & 0 & 0 & 0 & 0 & 0\\
 & 1 & 0 & 0 & 0 & 0 & 0\\
 & 1 & 0 & 0 & 0 & 0 & 0\\
 & 1 & 0 & 0 & 0 & 0 & 0\\ \hline
 	\end{tabular}
	\label{tab:T1}
\end{table}

We followed a similar process to produce a synthetic mechatronic software system comprised of the physical product architecture network of the Pratt Whitney (PW) aircraft engine\cite{Sosa2003} and synthetic software networks. Analysis of this system allows us to understand how the principles of failure tolerance of mechatronic software systems might apply in a real-world system.

\subsection{Topological attacks}
Following a general method in complex networks to study the failure tolerance of networks, we `attack' the hardware-software network in one of four ways: random, degree centrality, betweenness centrality and closeness centrality based attacks. The `attack' of a node models a hardware or software component no longer behaving within its performance envelope, i.e., component performance satisfies its marginal value~\cite{Youn2011}. Random attack signifies the failure of components due to gradual degradation. The last three types of attack are based on the premise of deliberately attacking the most significant nodes in the network. All these measures characterise the importance of the node to the network, on the basis that the more connected a node is, the more important the node is to the network, and that critical subsystems share important connections. As there is not necessarily one correct metric of node centrality, we investigate three common ones to determine which is the most conservative in relation to attack tolerance. Degree centrality is based on the degree or the number of links that a particular node has. The betweenness centrality is a measure of the number of paths that pass through a particular node. Closeness centrality measures how close a particular node to the other nodes in the network. Thus, the centrality attacks provide three mechanisms to simulate the failure of important components. In each of the centrality based attacks, the node with the highest centrality value is removed from the network in an iterative manner, until the network becomes fully disintegrated. In contrast, the removal of a randomly selected node in each iteration emulates the random failure of components. When simulating all these attacks, we make the assumption that the component failures occur gradually and incrementally; thus, we do not take into account the instances where more than one component would fail simultaneously. 

The betweenness centrality and the closeness centrality are calculated based on well-known algorithms~\cite{Sabidussi1966}. Eqn.~(\ref{eq:betweenness}) gives the formal definition of betweenness centrality.

\begin{equation}
BC(v) = \frac{1}{(N-1) (N-2)} \sum_{s \neq v \neq t} \frac{\sigma_{s,t}(v)}{\sigma_{s,t}}
\label{eq:betweenness}
\end{equation}

Here, $\sigma_{s,t}$ is the number of shortest paths between the source node $s$ and the target node $t$. $\sigma_{s,t}(v)$ is the number of shortest paths between source node $s$ and target node $t$ that lies through node $v$.

\begin{equation}
CC(v) = \frac{1}{\sum_{i \neq v} d_{g}(v,i)}
\label{eq:closeness}
\end{equation}

Eqn.~(\ref{eq:closeness}) defines how the closeness centrality~\cite{Sabidussi1966} of a node is measured. Here, $d_{g}(v,i)$ denotes the shortest path (geodesic) distance between nodes $v$ and $i$. The average is inverted so that the node that is closest to the other nodes will have the highest closeness centrality.

Degree centrality is directly related to the degree of each node in the network. Thus, $DC(v) = deg(v)$.

\subsection{Topological robustness and robustness coefficient}

The ability of a network to sustain or withstand random failures or targeted attacks depends on its topological structure. For example, scale-free networks have been shown to be more resilient against random failures, but are more vulnerable to targeted centrality based attacks, in comparison to Erd\"{o}s-R\'{e}nyi random networks\cite{Albert2000}. Thus, it is important to observe the topological robustness of a network to comprehend how its topological structure would contribute to random node failures or targeted attacks. 


There exists a substantial body of work which introduces and analyses structural robustness measures. Albert et al.~\cite{err-2000} considered error and attack tolerance of complex networks by comparing the profiles of quantities such as the Network diameter and the size of the largest component. Following their work, a multitude of metrics have been proposed to measure the topological robustness of networks as a single quantity. However, they typically  calculate averaged effects of single node removals, rather than effects of sequential removals, or are too simplistic. For example, the \textit{network efficiency} has been defined as the average of inverted shortest path lengths~\cite{err-2004}, and used for quantifying the robustness of a network. Node removals are not explicitly considered in this measure. Similarly, Dekker and Colbert~\cite{dstorobust} introduced two concepts of connectivity for a graph which can be used to model network robustness: the \textit{node connectivity } and \textit{link connectivity}, which are the smallest number of nodes and links respectively, whose removal results in a disconnected or single-node graph. In this work, we used robustness coefficient\cite{Piraveenan2012} as the robustness measure as it has the advantage of providing a single numeric value to quantify the topological robustness of a network under sustained attack.


   
The calculation of the robustness coefficient utilises the size of the largest connected component in a network. When a network is `attacked', or when a particular node is removed, this could alter the size of the largest connected component (a group of connected nodes) of the network. For each network, the number of nodes that has to be removed to disintegrate the network completely would be different. Thus, if the size of the largest connected component is plotted against the number of nodes removed, that would give an indication on how resilient the network is against targeted attacks and random failures. The topological metric provides a proxy for system performance without having to calculate the actual system performance, which can be computationally expensive unless methods such as response surface are applied to estimate performance~\cite{kim:121401}. As the robustness coefficient decreases, more of the system is behaving outside of its marginal value. 

\begin{figure}[t]
\includegraphics[width=9.05cm]{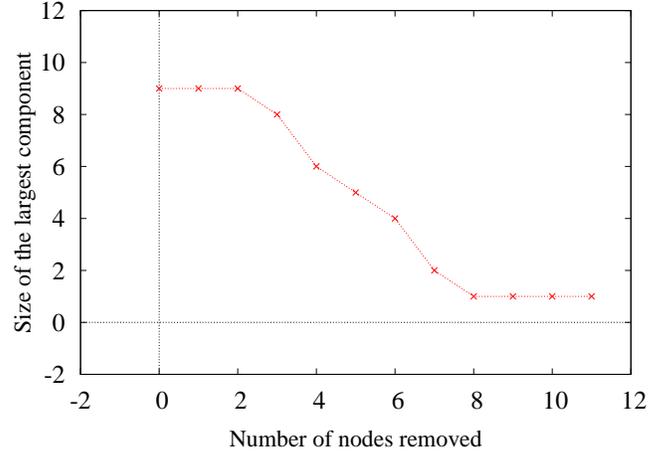}
\caption{\label{fig:targetedattack}SIZE OF LARGEST COMPONENT AGAINST THE NUMBER OF NODES REMOVED FOR A NETWORK UNDER SUSTAINED TARGETED ATTACK}
\end{figure}

\begin{figure}[t]
\includegraphics[width=9.05cm]{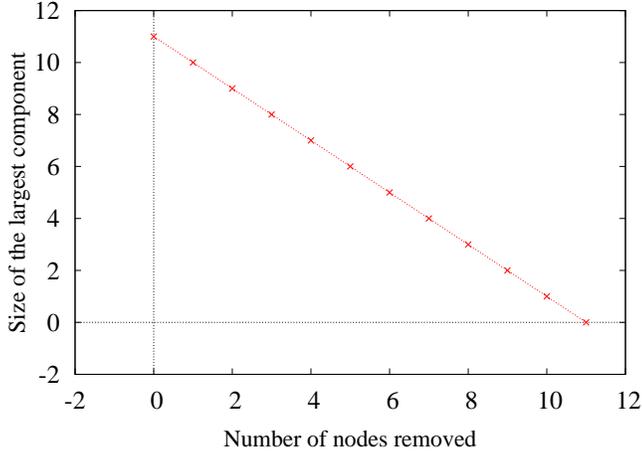}
\caption{\label{fig:targetedattackideal}SIZE OF LARGEST COMPONENT AGAINST THE NUMBER OF NODES REMOVED FOR AN IDEALLY ROBUST NETWORK UNDER TARGETED ATTACK}
\end{figure}

Fig. \ref{fig:targetedattack} shows a typical curve that is generated by plotting the size of the largest connected component against the number of nodes removed. The area under this curve can be regarded as a measure of attack tolerance or the topological robustness of a network under sustained attack. By accumulating the areas of trapeziums of unitary width along the x-axis, the area under the curve could be calculated as,

 \begin{align*}
\label{eqD0}A_1&= 0.5(S_0+S_1)+0.5(S_1+S_2)+......+0.5(S_{N-1}+S_N) \\
&= 0.5 S_0 + {\sum \limits_{k=1}^{N-1} {S_k} } + 0.5 S_N
\end{align*} where $S_k$ is the size of the largest component after k nodes are removed.

Here $S_0$ is the initial largest component size. Since $S_N$, the size of the largest component after N nodes are removed, is by definition zero, the above expression can be reduced to Eqn.~(\ref{eqD1}). 

\begin{equation}
\label{eqD1}A_1=  {\sum \limits_{k=0}^{N} {S_k} } - 0.5 S_0
\end{equation}

Fig. \ref{fig:targetedattackideal} shows the ideal curve that could occur in a network under attack. In an ideal scenario, with the removal of each node, the size of the largest connected component would get reduced in a linear fashion. The size of the largest component would always decrease only by 1, in each iteration. Since the particular curve forms a triangle of base $N$ and height $N$, the area under it can be calculated by, 

\begin{equation}
\label{eqD2}A_2= (1/2)N^2
\end{equation} 

The ratio between the area under the equivalent graph of any network and the area under the ideal curve would give a measure of the topological robustness of the network of concern. This measure is called the `robustness coefficient' of a network. Thus, the robustness coefficient $R$ would be, 

\begin{equation}
\label{eqD7}{R}=\frac{A_1}{A_2}= \frac{2 {\sum\limits_{k=0}^{N} {S_k}  - S_0}}{N^2}
\end{equation}

Eqn.~(\ref{eqD8}) represents $R$ as a percentage. 

\begin{equation}
\label{eqD8}{R}= \frac{A_1}{A_2}= \frac{200 {\sum\limits_{k=0}^{N} {S_k}  - 100S_0}}{N^2}
\end{equation}

Therefore, the formal definition of the robustness coefficient can be given as,

\begin{equation}
R = \frac{200 \sum_{k=0}^{N} S_{k} - 100 S_{0}}{N^{2}}
\label{eq:robustness}
\end{equation}

In Eqn.~(\ref{eq:robustness}), $S_{k}$ is the size of the largest component after $k$ nodes are removed. $S_{0}$ denotes the initial largest component size. $N$ is the network size. According to the above calculation, for a fully connected network of any size, the robustness coefficient ($R$) would always be 100\%.

In order to measure the robustness coefficient under different types of attacks, the nodes that are removed are selected based on their centrality values. For example, in a betweenness based attack, the node that would be removed in each iteration would be the node with the highest betweenness centrality value, at that particular instance of the network. Likewise, the other centrality based attacks would select the node with the highest value of the respective centrality measure. In the case of a random attack, a randomly selected node would be removed in each iteration.

The interactions between the hardware and software networks in a mechatronic system form a separate network among the hardware components and their respective software drivers. The total network that is formed by the hardware component architecture network, software network and the interactions of these two networks provide an interesting platform on which the topological robustness measures such as the robustness coefficient could be applied. When the hardware-software networks evolve over time, due to the changes in requirements or the addition of new features into the system, the total network topology of the original system may become altered in an unexpected manner. How these unexpected topological changes affect the topological robustness of the system may not be obvious unless a topological robustness analysis of the total system is performed. Therefore, it is very important to observe the topological robustness of hardware-software networks using measures such as the robustness coefficient in order to better understand how the robustness of the overall system is affected by its topological changes. 

Engineering robustness of a system is its ability to perform in an acceptable manner under the expected variations of certain parameters, but also in the presence of unexpected variations in other unknown parameters. However, the topological robustness of a network focuses on sustaining the connectivity of the nodes under random failures or deliberate sustained attacks. Therefore, engineering robustness can be thought of as a measure of a system's ability to maintain its functionality under random failures or targeted attacks. On the other hand the topological robustness is a measure of a system's ability to maintain the connectivity among its components under random failures or targeted attacks from a network analysis point of view. Even though the topological connectivity may affect a system's functionality, it is important to note that there exists a significant difference between these two perspectives of robustness.

\section{Results}
\label{results}
\subsection{Topological robustness of hardware networks}
As the initial part of the research, a comparison of topological robustness between modular and hierarchical modular architectures was performed. Table \ref{tab:robustmodular} shows the results obtained from that comparison. The hierarchical modular networks were generated using varying rewiring probability ($p$) values to change the modularity in each network. Each of these networks tested had 100 nodes with 5 modules.

\begin{table}[htbp]
	\centering
		 \small		
			 \caption{ROBUSTNESS COEFFICIENT VALUE COMPARISON OF MODULAR AND HIERARCHICAL MODULAR NETWORKS}
\begin{tabular}{lrrrr}
	\hline
	Type of Attack & \multicolumn{4}{c}{Robustness Coefficient (\%)} \\
	\hline
 & $p=0$ & $p=0.2$ & $p=0.5$ & $p=0.8$\\

 Degree based & 22 & 93 & 91 & 91\\
 Betweenness & 22 & 83 & 88 & 88\\
 Closeness & 36 & 91 & 94 & 93 \\
 Random & 24 & 98 & 99 & 97\\ \hline
 	\end{tabular}
	\label{tab:robustmodular}
\end{table}

In the figures given in Table \ref{tab:robustmodular}, the network generated with $p=0$ represents a perfectly modular network and an increasing $p$ indicates increasing degree of hierarchical modularity. According to these results it is evident that the hierarchical modular networks show higher degree of topological robustness compared to modular networks under sustained attacks because the failures can be isolated within a nested module. It is important to note that we are referring to the `topological robustness' of a network in contrast to the engineering robustness that may be more directly applicable to engineering systems. Generally, engineering systems with higher modularity are regarded as possessing higher engineering robustness, but only when the failure can be isolated to a single module. Nevertheless, these results provide an interesting baseline on topological robustness of an engineering system from a network analysis point of view.

We performed a topological robustness analysis on the actual hardware component network of the Pratt Whitney Aircraft Engine component network~\cite{Sosa2003}. Since this will be the hardware network that we will subsequently use as part of our hardware-software integrated network, it was necessary to observe its own robustness coefficient measures first. Table \ref{tab:pwrobustness} shows the robustness coefficient results obtained for the Pratt Whitney (PW) Aircraft Engine component network~\cite{Sosa2003} under different types of sustained attacks.

\begin{table}[htbp]
	\centering
		 \small		
			 \caption{ROBUSTNESS COEFFICIENT VALUES FOR THE PW AERO-ENGINE UNDER DIFFERENT TYPES OF ATTACKS}
\begin{tabular}{lr}
	\hline
	Type of Attack & Robustness Coefficient (\%) \\
	\hline
 Degree based & 85\\
 Betweenness & 68\\
 Closeness & 79\\
 Random & 91\\ \hline
 	\end{tabular}
	\label{tab:pwrobustness}
\end{table}

According to the above results, the PW aircraft engine network shows higher robustness against random node failures, compared to the centrality based attacks.

\subsection{Topological robustness of software networks}
In order to evaluate a hardware-software network's behaviour under different kinds of attacks, we generated three hypothetical software networks of varying sizes. The sizes were selected in such a way that they would have approximately 1:2, 1:5 and 1:10 ratios (with the values 100, 250 and 500) to the size of the PW engine network (54). The isolated nodes were pruned from the generated networks, which reduced their sizes from the projected values.

Software networks have been demonstrated to have a scale-free architecture~\cite{Valverde2002}. Therefore, the software networks were generated using preferential attachment, since preferential attachment model is known to generate scale free networks~\cite{Barabasi1999}. Table \ref{tab:softwarerobustness} shows the robustness coefficient values of three scale-free software networks that were generated, under different types of attacks.

\begin{table}[htbp]
	\centering
		 \small		
			 \caption{ROBUSTNESS COEFFICIENT OF SCALE-FREE SOFTWARE NETWORKS OF DIFFERENT SIZES (N) UNDER DIFFERENT TYPES OF ATTACK}
\begin{tabular}{lrrr}
	\hline
	Type of Attack & \multicolumn{3}{c}{Robustness Coefficient (\%)}\\
	\hline	
	& $N=95$ & $N=233$ & $N=470$\\
Degree & 26 & 23 & 26\\
Betweenness & 23 & 21 & 23\\
Closeness & 39 & 38 & 39\\
Random & 73 & 71 & 73\\ \hline
 	\end{tabular}
	\label{tab:softwarerobustness}
\end{table}

All three hypothetical software networks demonstrate higher topological robustness against random attacks, compared to centrality based attacks. This is comparable with the results obtained for the PW engine network, but differs from the modular and hierarchical modular hardware networks. However, in the software networks considered, the margin of difference between topological robustness under random failures and centrality based attacks is much higher than that of the PW engine network.

\subsection{Topological robustness analysis of hardware-software networks}

We now test the integration of the hardware and software networks. We start with a random integration of the modular and hierarchical modular hardware networks with a scale-free software network, that is, randomly setting the adjacency values in C and D of Fig. \ref{fig:F1}.

\begin{table}[htbp]
  \centering
\small 
  \caption{ROBUSTNESS COEFFICIENTS OF RANDOMLY INTEGRATED HW-SW INTEGRATED NETWORKS OF MODULAR AND HIERARCHICAL MODULAR HARDWARE NETWORKS AND A SCALE-FREE SOFTWARE NETWORK UNDER DIFFERENT TOPOLOGICAL ATTACKS}
    \begin{tabular}{rrrrrrr}
\hline 
    Type of Attack & \multicolumn{3}{c}{Modular HW--SW} & \multicolumn{3}{c}{HM HW--SW} \\
   \hline 
          & 10\% & 20\% & 50\% & 10\% & 20\% & 50\% \\
    Degree & \multicolumn{1}{c}{47} & \multicolumn{1}{c}{47} & \multicolumn{1}{c}{47} & \multicolumn{1}{c}{47} & \multicolumn{1}{c}{47} & \multicolumn{1}{c}{47} \\
    Closeness  & \multicolumn{1}{c}{59} & \multicolumn{1}{c}{60} & \multicolumn{1}{c}{60} & \multicolumn{1}{c}{58} & \multicolumn{1}{c}{59} & \multicolumn{1}{c}{60} \\
    Betweenness & \multicolumn{1}{c}{45} & \multicolumn{1}{c}{47} & \multicolumn{1}{c}{45} & \multicolumn{1}{c}{45} & \multicolumn{1}{c}{45} & \multicolumn{1}{c}{45} \\
    Random & \multicolumn{1}{c}{98} & \multicolumn{1}{c}{98} & \multicolumn{1}{c}{99} & \multicolumn{1}{c}{97} & \multicolumn{1}{c}{99} & \multicolumn{1}{c}{99} \\
\hline 
    \end{tabular}%
  \label{tab:hmmodsw}%
\end{table}%

Table \ref{tab:hmmodsw} shows the robustness coefficient values of randomly integrated modular and hierarchical modular synthetic hardware networks with a synthetic scale-free software network. The hardware networks that were used had 100 nodes each, containing exactly 5 modules. The rewiring probability $p$ used for generating the hierarchical modular network was 0.5. The software network that was used had 470 nodes. Three different adjacency probabilities (10, 20 and 50\%) were considered when integrating the networks.

The results do not show a significant difference between the two different types of integrated networks, that is, no advantage for hierarchical modular network. In contrast, the modular and hierarchical modular hardware networks demonstrated a significant difference in topological robustness in favour of hierarchical modular as per the results shown in the table \ref{tab:robustmodular}. Thus, it is possible to argue that even though the hardware networks of different modularity levels may show varying topological robustness characteristics, when they are integrated with scale-free software networks, the topological robustness of the resulting integrated networks tend to be uniform in nature. It may be that scale-free software network dominates the failure characteristic, since the scale-free software network is vulnerable to targeted attacks as shown in Table \ref{tab:softwarerobustness}. Apart from that, we can observe that the variation of the number of interconnections (denoted by the different percentages of adjacency levels) do not have a significant impact on the robustness coefficient values. This implies that the redundancy of random interconnections between the hardware and software networks does not necessarily improve the topological robustness of the resulting integrated network.

Next, we measured the robustness coefficients of several integrated hardware-software networks consisting of modular and hierarchical modular synthetic hardware networks and a synthetic scale-free software network. These networks were integrated according to several well-known motifs. A randomly integrated network of a similar adjacency levels was used for comparison. Table \ref{tab:hm_mod_sw_motif} shows the results obtained from the measurements. The hardware networks had 100 nodes and 5 modules each. The hierarchical modular network was generated with a rewiring probability $p$ of 0.5. The scale-free software network used had 470 nodes. 

\begin{table}[htbp]
  \centering
\small 
  \caption{ROBUSTNESS COEFFICIENTS OF HW-SW INTEGRATED NETWORKS OF MODULAR AND HIERARCHICAL MODULAR HARDWARE NETWORKS AND A SCALE-FREE SOFTWARE NETWORK, INTEGRATED WITH HETEROGENEOUS ARCHITECTURES, UNDER DIFFERENT TOPOLOGICAL ATTACKS}
    \begin{tabular}{rcccc}
\hline 
    Type of Attack & \multicolumn{4}{c}{Modular HW--SW} \\
    \hline 
          & Bus & Ring & Star & Random \\
    Degree & 38    & 32    & 39    & 38 \\
    Closeness  & 47    & 55    & 47    & 50 \\
    Betweenness & 23    & 15    & 22    & 29 \\
    Random & 75    & 76    & 76    & 80 \\
\hline 
    Type of Attack & \multicolumn{4}{c}{HM HW--SW} \\
\hline 
          & Bus  & Ring & Star & Random \\

    Degree & 38    & 39    & 31    & 39 \\
    Closeness  & 52    & 53    & 52    & 51 \\
    Betweenness & 25    & 25    & 17    & 31 \\
    Random & 78    & 83    & 77    & 78 \\
\hline 
    \end{tabular}%
  \label{tab:hm_mod_sw_motif}%
\end{table}%

The robustness coefficient values given in the Table \ref{tab:hm_mod_sw_motif} show that when a hierarchical modular hardware network is integrated with a scale-free software network, bus and ring architectures would generally introduce more topological robustness compared to the star architecture. Apart from that, the random integration gives similar robustness coefficient values, implying that following a specific architecture when integrating the hardware and software networks does not significantly improve the resulting network's topological robustness. This would indicate that the ``designed" system architecture does not show significantly higher robustness than the random integration of comparable sized networks. Also, when a hierarchical modular hardware network is used, the ring and bus architectures would not substantially improve the topological robustness of the resulting integrated network.  

Finally, we integrate the PW engine hardware network with the hypothetical scale-free software networks. First we tested random integration. This was done by randomly populating the hardware-software adjacency matrix sections C and D of Fig. \ref{fig:F1}. Three different percentages (10, 20 and 50\%) of probability of connection in the adjacency matrix sections were considered. Then, the robustness coefficients of the resulting integrated networks were measured. Table \ref{tab:pwsoftwarerobust} shows the results of those measurements.

\begin{table}[htbp]
	\centering
		 \small		
			 \caption{ROBUSTNESS COEFFICIENTS OF HW-SW INTEGRATED NETWORKS OF VARYING RANDOM ADJACENCY LEVELS, UNDER DIFFERENT TYPES OF ATTACKS}
\begin{tabular}{lrrr}
	\hline
	Type of Attack & \multicolumn{3}{c}{HW ($N=54$) SW ($N=95$)}\\
\hline 
      & 10\% & 20\% & 50\%\\
Betweenness Attack & 65 & 68 & 68\\
Closeness Attack & 74 & 80 & 75\\
Degree Attack & 67 & 69 & 70\\
Random Attack & 95 & 98 & 99\\
	\hline
	Type of Attack  & \multicolumn{3}{c}{HW ($N=54$) SW ($N=233$)}\\	
\hline 
	& 10\% & 20\% & 50\%\\
Betweenness Attack & 47 & 48 & 48\\
Closeness Attack & 59 & 58 & 59\\
Degree Attack & 49 & 50 & 50\\
Random Attack & 95 & 98 & 99\\
	\hline
	Type of Attack & \multicolumn{3}{c}{HW ($N=54$) SW ($N=470$)}\\	
\hline 
	 & 10\% & 20\% & 50\%\\
Betweenness Attack & 35 & 35 & 35\\
Closeness Attack & 50 & 51 & 52\\
Degree Attack & 37 & 37 & 37\\
Random Attack & 96 & 98 & 99\\
	\hline
 	\end{tabular}
	\label{tab:pwsoftwarerobust}
\end{table}

All the integrated networks show higher topological robustness against random failures compared to targeted centrality based attacks. Another important observation that can be made is that when the software network used increases in size, the respective hardware-software network's topological robustness against centrality based attacks tends to decrease. In other words, as the amount of code increases, the overall system tolerance to failure decreases. However, the software network size does not seem to affect the topological robustness of the integrated network against random failures. 

Another key observation that could be made out of this comparison is that the increase in the adjacency level does not seem to affect the topological robustness of the integrated network in a substantial manner. Though more connections among the hardware and software layers increase redundancy, it does not seem to positively affect the topological robustness of the hardware-software network.
  
Finally, we combined the PW hardware network and the software networks according to different well-known architectures by following the adjacency matrix patterns given in Table \ref{tab:T1}. Table \ref{tab:motifrobust} shows the robustness coefficient values of the hardware-software networks that were obtained under different architectures, against different types of attacks. The software network used has a network of node count 233. A randomly integrated network with a similar adjacency level was also used for the comparison.

\begin{table}[htbp]
	\centering
		 \small		
			 \caption{ROBUSTNESS COEFFICIENTS OF HW-SW NETWORKS OF HETEREOGENEOUS ARCHITECTURES, UNDER DIFFERENT TYPES OF ATTACKS}
\begin{tabular}{lcccc}
	\hline
Type of Attack & Ring & Bus & Star & Random\\
\hline 
Betweenness Attack & 27 & 27 & 17 & 32\\
Closeness Attack & 50 & 46 & 48 & 50\\
Degree Attack & 37 & 37 & 27 & 38\\
Random Attack & 79 & 75 & 77 & 78\\
	\hline
 	\end{tabular}
	\label{tab:motifrobust}
\end{table}

Based on the results given in Table~\ref{tab:motifrobust}, it is possible to argue that the ring and bus architectures show better topological robustness against centrality based attacks, compared to the star architecture. Another interesting observation that can be made is how the randomly integrated network manages to demonstrate comparable topological robustness, under all types of attacks. This may suggest that following a particular architecture in hardware-software integration may not result in a substantial improvement in the topological robustness of the integrated network.

\section{Discussion and Conclusion}
This paper introduced a complex networks approach to calculate the failure tolerance of mechatronic software systems to the failure of critical hardware components and software code (functions and classes). We used a novel approach in modelling the integration of interdependent hardware component networks and software networks in mechatronic software systems, according to different system architectures. Calculation of the robustness of the system is based on a robustness coefficient, which is a topological measure of the size and connectedness of the network. We investigated the robustness of several synthetic hardware-software networks comprised of modular and hierarchical modular product architectures and scale-free software architectures connected randomly or through known connection architectures and a synthetic network comprised of a Pratt and Whitney aero-engine and a scale-free software network. In all instances, the results show that random integration of the product architecture to the software code results in comparably resilient hardware-software systems compared to the designed counterparts of star, bus, and ring architectures. Although actual engineered systems are not randomly integrated, this result implies that the attempts to optimise system connectivity through certain architectures may not necessarily make the system more resilient to topological failure. The challenge of system engineering is to balance the assembly efficiency gains introduced by prescribed interconnections between components and the possible robustness gains introduced by directly connecting components to their interacting counterparts without any prescribed regularity. This is what is meant by a `random' connection architecture, not that components are randomly connected to other components.

Centrality-based attacks produced the most failure in the hardware-software system (most significant reduction in robustness coefficient), with betweenness centrality attacks producing the most significant effects. The scale-free architecture of software networks strongly determined the vulnerability of the hardware-software system, thus highlighting the importance of producing error-free code but also bringing into question the suitability of software code with highly dependent functions and classes in mission-critical systems.

Future work will include correlating the robustness coefficient with the actual performance of the system to determine more accurately whether the robustness coefficient could serve as a useful and efficient proxy for system performance, given the computational complexity involved in simulating the performance of a complex system. As well, we will investigate new design methods for `wiring up' product architectures to their software controllers with the aim of improving the overall robustness of the system as a first-class concern. The complex networks approach applied in this research can be used for future work in understanding the failure tolerance of complex engineered systems with interdependent networks.

Although the scope of the paper is focused on mechatronic systems, it maybe generalised for complex engineered systems with inter-dependent networks. However, in this work, we opted to focus more on mechatronic software systems as the modelling formalism used relates more to such systems. Generalising this particular modelling approach for complex engineered systems would be a possible extension to this research.

\label{conclusion}

\bibliographystyle{asmems4}

\begin{acknowledgment}
Andy Dong is the recipient of an Australian Research Council Future Fellowship (project number FT100100376).
\end{acknowledgment}

\bibliography{dtm2013_attackresilience}



\begin{acronym}[TDMA]
\acro{MFIP}{model-based failure identification and propagation}
\acro{FFIP}{Functional Failure Identification and Propagation}
\end{acronym}

\end{document}